# Mode-locking based on a zero-area pulse formation in a laser with a coherent absorber

MIKHAIL V. ARKHIPOV,[1] ALEXANDER A. SHIMKO,[1] ALEXEY A. KALINICHEV[1], IHAR BABUSKIN[3,4], NIKOLAI N. ROSANOV[2,5], AND ROSTISLAV M. ARKHIPOV[1,2*]

[1]Faculty of Physics, St. Petersburg State University, Ulyanovskaya St. 1, Petrodvoretz, St. Petersburg, 198504, Russia
[2] ITMO University, Kronverksky Pr. 49, St. Petersburg, 197101, Russia
[3] Institute of Quantum Optics, Leibniz University Hannover, Welfgarten 1, 30167 Hannover, Germany
[4] Max Born Institute, Max Born Str. 2a, 12489 Berlin, Germany
[5] Vavilov State Optical Institute, Kadetskaya linia 5,199053 St. Petersburg, Russia

*Corresponding author: arkhipovrostislav@gmail.com

**We observe experimentally a mode-locking in a continuous narrow-band tunable dye laser with molecular iodine absorber cells, which transitions have large phase relaxation time $T_2$. We show that the mode-locking arises due to coherent interaction of light with the absorbing medium leading to Rabi oscillations, so that zero-area ($0\pi$-) pulses in the absorber are formed. Such mode-locking regime is different to most typical passive modelocking mechanisms where saturation plays the main role.**

______________________________________________

Mode locked (ML) lasers are principal sources of short light pulses with a high repetition rate [1, 2]. The most commonly used method to create such pulses inside a cavity is a passive mode-locking. Typically such lasers consist of two sections, one with a gain medium and the other with a nonlinearly absorbing one. The physical mechanism of mode-locking in such lasers is based on the effect of saturation of absorption in the absorber section as well as the saturation of the gain in the amplifier.

Recently, another mechanism of passive mode-locking was proposed, where the absorber and gain media operate not in the regime of saturation but in the coherent regime of strong coupling of light with the medium [3-10]. Coherent light-matter interaction regime arises when light pulse duration is smaller than medium polarization relaxation time $T_2$ [11-13]. The presence of phase memory leads to the novel unusual phenomena in the propagation of the pulse in nonlinear absorbing or amplifying media, which cannot be observed when the regime of light-matter interaction is incoherent [3-25]. For example, pulse can propagates without losses in the regime of self-induced transparency (SIT) in the absorber as a $2\pi$ pulse [11]. In the amplifying medium stable $\pi$ pulse can exist [12]. The key idea of the mode-locking based on the coherent effects is to use a configuration allowing the same pulse to be a $2\pi$-pulse in the absorber and $\pi$-pulse in the amplifier [3-10]. In this way, since a $2\pi$-pulse is a stable soliton solution, absorber dumps any background perturbations making the regime stable. At the same time, a $\pi$-pulse configuration in the amplifier allows to take the energy from it, thus leading to a pulse shortening.

Despite of the great potential promising to deliver pulses up to single-cycle duration [6, 7], lasers with such coherent mode-locking mechanism have not been realized experimentally yet. One of the problems is to obtain coherent interaction inside the cavity for the parameters required for the mode-locking. Although coherent effects were observed in plenty of optical systems [11-26], achieving intra-cavity coherent effects combined with a mode-locking is not a trivial task. In particular, they were observed in single-section gas lasers [16], quantum dot amplifiers [17] and quantum cascade lasers in the THz frequency range [18]. Furthermore, in the recent paper [19], the signatures of SIT regime in a rubidium [78]Rb gas cell placed in a cavity of a mode-locked laser were demonstrated experimentally. However, the SIT regime in [19] was not related to the mode-locking. In the opposite, the quality of the SESAM- induced mode-locking degraded in the presence of the SIT. Despite of this, the experiment in [19] clearly demonstrated the feasibility of SIT inside the laser cavity. On the other hand, a possibility of the coherent regime onset in SESAM absorbers were predicted theoretically [27] but never tested experimentally.

Besides, a SIT pulse is not the only pulse shape, which propagates without losses in absorbing resonant media. Another possibility are so called zero-area pulses ($0\pi$-pulses) [20-23, which were recently theoretically shown to be also useful in the coherent mode-locking [10].

In this letter, we demonstrate a possibility of a passive mode-locking in a narrow-band dye laser when absorbing cells with molecular iodine vapors are placed in the cavity. As was preliminary reported in [24], since molecular iodine vapors have very large coherence time $T_2$ in the range of hundreds of nanoseconds [26], the absorber should be in the coherent regime. Here we provide solid experimental evidences that this is indeed the case and that the coherent interaction with the iodine cell causes the mode-locking. Besides, our experimental results corroborated with numerical simulations show that zero area pulses (0π-pulses) are formed.

The choice of molecular iodine vapor at room temperature as a candidate for a coherent absorber in our experiments is not accidental. It is a common object for observation of coherent interactions in the visible wavelength range such as photon echo, transient nutations, etc. [26]. Certain disadvantage of molecular iodine for our experiments is a high density of absorption lines in the generation region of the rhodamine 6G laser. Nevertheless, this allowed us to observe the mode-locking effect on various absorption lines.

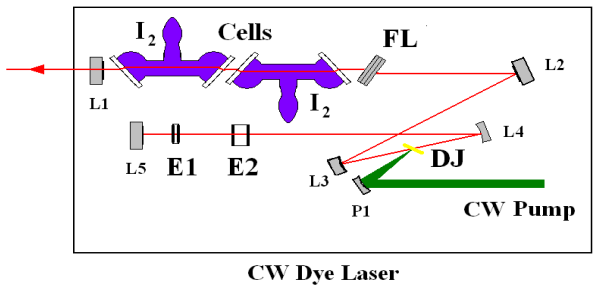

Fig. 1. Experimental setup of the narrowband dye laser used in our experiments. A dye jet DJ in the cavity formed by mirrors L1-L5 is pumped by a CW laser. A Lyot filter FL and two Fabry-Perot etalons E1, E2 were used to narrow the generation line. In addition, two $I_2$ absorber cells were placed in and outside the cavity. We used several variants of the setup, different by the presence or absence of the cells and selective elements in the cavity.

As a basis for our experimental setup we used a narrowband CW dye laser as shown in—Fig.1. This continuous dye laser was assembled on the basis of a laser module "T & D scan laser system" from TekhnoScan Company. The pump was produced by a laser "Verdi V10" firms Coherent. A linear 97 cm-long cavity of the dye laser was defined by five mirrors L1-L5. As a dye jet (DJ) an ethylene glycol solution of the dye Rhodamine 6G was used. It was placed between the two spherical mirrors L3 and L4, and has been deployed in a horizontal plane at the Brewster angle relative to the direction of lasing. The pump radiation was focused into the dye jet using a mirror P1. To narrow the generation line a Lyot filter (LF) and Fabry-Perot etalons (E1) and (E2) were used. The thickness of the first etalon was fixed to 0.3 mm. The second etalon E2 was taken in the form of plane-parallel glass plates with the thicknesses varying from 1 mm to 5 mm. In particular, when only the Lyot filter was placed in the cavity without the etalons, the lasing bandwidth was about 20-30 picometers. Simultaneous placement of the FL and E1 led to the linewidth less than 5 picometers.

We also used two cells with molecular iodine $I_2$ to place them inside or outside of the cavity. The cells windows were deployed at the Brewster angle.

The laser intensity was registered with a time resolution of about 0.5 ns provided high-speed photodiode 11HSP-V2 model and digital oscilloscope Agilent DSO9104A. The registration system contained also a power meter MAESTRO (by firm Standa) and spectrometer of laser control module produced by TekhnoScan Company that controlled the operation of stepping motors associated with intracavity filters. The spectra were registered using a Fabry-Perot interferometer. The interferograms of the emission spectra were recorded by a digital camera.

We start our consideration from the case when no selective elements and no cells are placed in the cavity. In this case, our laser setup generated a broad spectrum and no pulsations of intensity were observed.

Second, we consider a configuration where a single iodine cell was placed in the laser cavity whereas the other cell was placed outside of the cavity, and no frequency selective elements were present in the setup. In this case, deeps appeared in the output spectrum whereas no mode-locking was observed. In both cells, no luminescence was observed, despite of considerably high output power. This is a clear indicator that the lasing frequencies do not coincide with the frequencies of the absorption lines in the cell in this case. Hence, in this configuration no interaction of the laser radiation with the resonance lines of the molecular iodine took place.

As the next step, a LF filter was placed into the laser cavity to narrow the spectrum. We tuned the LF filter to the spectral range between 585.0 and 585.3 nm. This range contained around 14 strong absorption lines as shown in Fig. 3 (blue line). In this modification of the setup, in contrast to the previous case, the observed output spectrum contained narrow absorption lines, which could be observed on the interferograms, as shown in Fig. 2 (lower panel). Nevertheless, no luminescence in the cells as well as no mode-locking were observed. Unstable self-pulsing similar to shown in Fig. 4a were observed. To ensure that the origin of these lines is the absorption in the cell and not the parasitic interference effects in the intracavity optical elements, we repeated experiments after cooling the cell. Cooling facilitates significant decrease of iodine pressure. To cool the cell we wrapped it in a wadding soaked with liquid nitrogen. Upon a cooling the narrow absorption lines disappear, and appear again when heating the cell back to the room temperature.

Finally, we further narrowed the output spectral line by placing the Fabry-Perot etalon E1 in the cavity. Inclination of the etalon allowed tuning the frequency across the maximum of transmission of the FL. In this case, away from the absorption lines unstable self-pulsing (but not mode-locking) regimes were observed similar to the ones for the previous case (Fig. 4a). However, when the laser frequency was tuned to the absorption lines of iodine, a mode-locking with long-term stability of pulse shapes appeared, see Fig4b. Mode-locking regimes were observed and well reproduced

at the following values of wavelengths: 580.020 nm, 585.136 nm, 585.155 nm, 585.179 nm, 585.181 nm, 585.220 nm. We observed that the generation line located at the position of the deep of the configuration without E1 and E2. (cf. see. Fig. 2, upper and lower panels). Also, in contrast to the previous variant of the setup, the mode-locking was always accompanied by a strong luminescence in both cells (the intracavity and the extracavity one). If we cool the intracavity cells in this setup, both mode-locking and luminescence in the intracavity cell disappears.

Appearance of a stable mode-locking regime on the frequencies corresponding to the resonances of the cell also shows clearly that the coherent interaction in the iodine cells plays the decisive role in the observed mode-locking mechanism.

We also measured the absorption in the $I_2$ cells in dependence on wavelength at the room temperature and compared it with the luminescence intensity in the cells (see Fig. 3). To test the absorption, the cells were illuminated by an independent tunable narrowband laser source that enabled to fix clearly the absorption regions, which, as one can see from Fig. 3, coincide with the luminescence peaks.

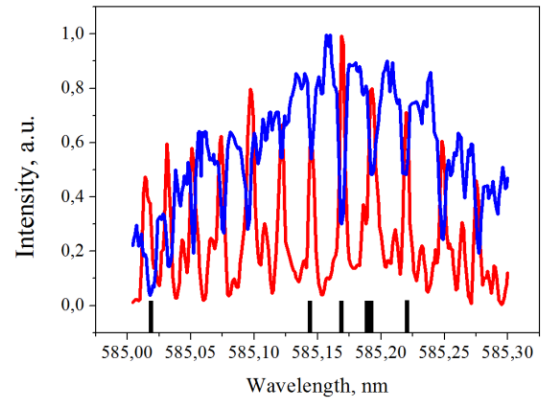

Fig. 3. The dependence of the intensity of radiation passing through the cell (blue line), and the luminescence intensity (red line) versus wavelength. The black lines indicate the positions where the mode-locking where observed.

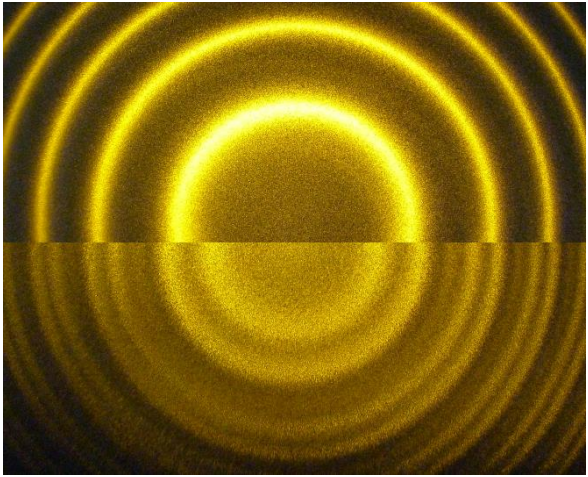

Fig.2. Example of interferograms of the laser spectrum with the iodine cells and FL in the cavity (lower panel) and with the iodine cells, FL, E1, in the cavity (upper panel).

Besides, within the resolution limit of our wavelengths meter they coincided also with the wavelengths at which it was possible to observe the mode-locking.

The later observation is very important, since sufficiently large absorption was observed in the cells at resonances. Because of this, one would expect that it should have led to a drop in power of generation in the mode-locking regime when the laser frequency is tuned to one of the cell absorption lines. However, our measurements showed that the output power does not have any notable drop near the absorption lines. This could only be the case, as already mentioned, when the passage of the pulses was not accompanied by a significant absorption.

To further clarify the role of absorption in molecular iodine additional experiments were carried out. As the spectrum of $I_2$ contains large number of absorption lines, one can suppose that the cells plays only an auxiliary role of "mode selectors" which introduce frequency-dependent losses and thus effectively narrowed the cavity resonance. To check if this possibility takes place we artificially narrowed the emission spectrum using the Fabry-Perot elements E1 and a thick glass plate E2, but without the cells in the cavity. In this case self-pulsating regimes were indeed observed in the cavity, see Fig. 4a. However, these regimes did not have long-term stability in contrast to the mode-locking regimes in the presence of the cells, see Fig. 4b.

The next important feature of the observed mode-locking was the double-peak pulse shape which is very typical for $0\pi$ pulses (see Fig. 5a). Besides, the relative intensities of the sup-peaks in the double-peak pulse experienced slow periodic modulation, which manifested themselves in the modulation of the envelope in Fig. 4b.

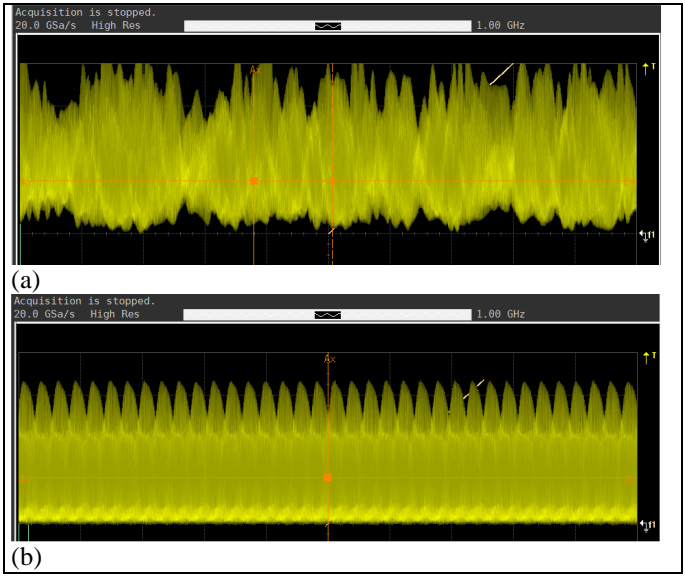

Fig. 4. The oscillograms of the output laser intensity measured for 20 μs. (a) A self-pulsing regime of the laser intensity without the cells and (b) a mode-locking regime in the case when the iodine cell were present in the cavity.

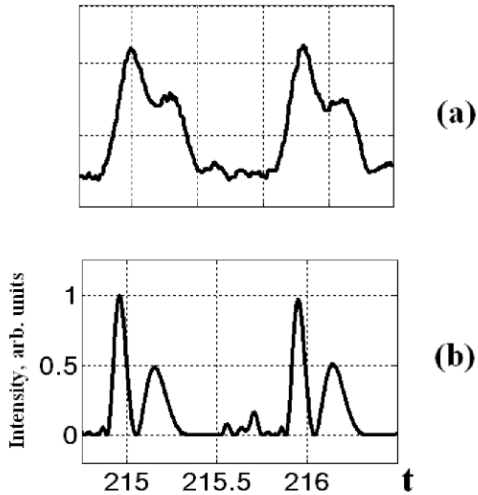

Fig.5. (a) An oscillogram (time scale 2.5 ns/div) of mode-locked pulses in the experimental setup. (b) The intensity of 0π-pulses according to the numerical simulations. The time t is normalized to the roundtrip time.

To corroborate our experimental findings we performed also numerical simulations, which showed rather good agreement with the experiment. For the numerical modeling a system of Maxwell-Bloch equations for the two-level medium described in [8-10] were used. Numerical simulations of these equations showed that the presence of the absorber with the dephasing time $T_2 > \tau$ ($\tau$- cavity roundtrip time) does not significantly influences the average output power of the laser and can only slightly reduce its value, which is in the agreement with the experimental results.

For the laser without the absorber section, due to the fact that the active medium occupies a small part of the cavity, we obtained either CW solutions or self-pulsating regimes.

Placing of a coherent absorber into the cavity led to significant modification of the dynamics. Pulses which were formed in the absorber had zero area (that is, they are 0π-pulses) and thus were passed through the absorber without losses. As it is usual for 0π-pulses, they store their energy in the medium and then retrieve it back with the phase shift π [20-22]. The typical example of a two-peak 0π-pulse obtained in our numerical simulations is shown in Fig.5b (intensity is presented). We also were able to reproduce experimentally observed slow pulsations dynamics mentioned above (not shown). This agreement with experiment supports both the coherent character of the mode-locking regime and the 0π-shape of the pulses in our experiment.

In conclusion, we observed mode-locking regimes in the cavity with the molecular iodine cell as an absorber. Despite of the complicated level structure of $I_2$ we were able to lock, using additional spectral filtering, the particular iodine lines. We have demonstrated that the mode-locking appears namely due to coherent character of the interaction of light with the iodine. The mode-locked pulses have zero area in the absorber. These findings were supported by numerical simulations, which are in a good agreement with our experiment.

**Funding.** Government of Russian Federation (074-U01); Russian Foundation for Basic Research (16-02-00762); German Research Foundation (DFG) (project BA 4156/4-1); Nieders. Vorab (project ZN3061).

**Acknowledgment.** The investigations were performed at the Center for Optical and Laser Materials Research, Research Park, St. Petersburg State University.